\documentclass[prl,twocolumn,showpacs,groupaddress]{revtex4-1}
\usepackage{graphicx,amsfonts,amsmath,amssymb}

\usepackage[toc,page]{appendix}
\usepackage{epsfig}

\usepackage{epstopdf}
\usepackage{times}
\usepackage{txfonts}

\begin{document}
\bibliographystyle{h-physrev3}
\title{Collision-induced frequency shifts in bright matter-wave solitons and soliton molecules}
\author{A. D. Martin}
\affiliation{Department of Physics, University of Otago, Dunedin}

\date{\today}
\begin{abstract}
A recent experiment has detected collision-induced frequency shifts in bright matter-wave solitons for the first time [Nat. Phys. 10, 918 (2014)]. Using a particle model, we derive the frequency shift for two solitons in a harmonic trap, and compare it to the results of the recent experiment and reported theoretical approximation. We find regimes where the frequency shift is much smaller than previously predicted, and propose experiments to test these findings.
We also predict that reducing the experimental trap frequency will reveal for the first time soliton molecules or soliton bound states in a cold atoms system. The dynamics of such bound states are found to be both highly phase-dependent, and sensitive to the residual 3D nature of the experiment. 
\end{abstract}

\pacs{ %change these
03.75.Lm,    %BEC inc. solitons
} \maketitle 

%\section{Introduction}
Since the first realisations of bright solitons in Bose-Einstein condensates (BECs) \cite{Strecker:2002, Khaykovich:2002, Cornish:2006}, many uses for solitons have been suggested including as surface-probes \cite{Lee:2006,Cornish:2009}, for Bell-state generation \cite{Gertjerenken:2013} and for interferometry \cite{Khawaja:2011,Martin:2012,Helm:2012}. The longstanding interest in bright solitons has recently been revived by an experiment on soliton collisions \cite{Nguyen:2014}. 

Solitons are solitary waves robust with respect to collisions; they survive these collisions unscathed up to a shift in their positions and phases. However, the vital characteristic of bright solitons -- the collisional shifts -- has not been observed in BEC until recently. The experiment of Rice University \cite{Nguyen:2014} of two solitons repeatedly colliding in a harmonic trap, demonstrated these shifts for the first time. The position shifts were revealed by their effect on the solitons' frequency of oscillation in the trap. The authors of Ref.\ \cite{Nguyen:2014} compared their results with an approximate theoretical curve, which agreed reasonably well with the experiment in the tested regime. 

In this Letter we provide an improved prediction of the frequency shift based on a particle model that reproduces the exact position shifts in the untrapped case \cite{Scharf:1992,Martin:2007, Martin:2008}. We identify regimes in which there is a measurable difference between our predictions and the theory of Ref.\ \cite{Nguyen:2014} and propose simple modifications to the experiment to explore these regimes. We also suggest modifications which will produce exotic soliton molecules (bound states) \cite{Khawaja:2011}. Optical soliton molecules have been produced experimentally \cite{Stratmann:2005}, but until now such states have not been produced in a cold atoms system. We predict that a reduction in the trap frequency of the experiment \cite{Nguyen:2014}, along with modifications to control the relative phase, would produce soliton molecules in BEC for the first time.
We investigate the 3-dimensional (3D) corrections to the dynamics, which affect both the stability and the mean-field dynamics of the BEC, particularly for the soliton molecules.

%\section{Theoretical Model}
Bose-Einstein condensates with tight radial trapping are in most circumstances well-described by the 1-dimensional (1D) Gross-Pitaevskii Equation (GPE) \cite{Dalfovo:1999}:
\begin{equation}
i\hbar \frac{\partial}{\partial t}\psi= -\frac{\hbar^2}{2m} \frac{\partial^2}{\partial x^2}\psi  + \frac{m\omega^2 x^2}{2}\psi+ g_{\textrm{1D}}N\left|\psi\right|^2\psi ,\label{Eq_GPE}
\end{equation}
where $g_{\textrm{1D}} = 2\hbar\omega_r a_s$, $N$ is the total atom number, $\omega$ and $\omega_r$ are the axial and radial trapping angular frequencies, and $m$ and $a_s$ are the mass and s-wave scattering length of the atomic species. In experiments, $a_s$ may be tuned to vary the sign and magnitude of the interactions, e.g., negative $a_s$ permits bright soliton solutions \cite{Billam:2012}.
We note that the 1D GPE breaks down in regimes where quantum or thermal fluctuations \cite{Proukakis:2008} and/or 3D effects become significant \cite{Parker:2008}. Quantum and thermal effects are not expected to play a large role in the situations considered in this paper. However, the slightly 3D nature of the geometry imposes a limit on the interaction strength, beyond which the condensate will collapse \cite{Nguyen:2014, Gammal:2001}. This limit is characterised by a critical atom number $|N_c|$. The parameter $N_c = 0.67 a_r/a_s$, where $a_r$ is the radial harmonic length of the trapping potential, is also used here and in Ref.\ \cite{Nguyen:2014} to characterise the strength of the interactions, since $N_s/N_c$ is proportional to $g_{1D}N$, and $N_s\approx N/2$ is the number of particles per soliton. Adding the quintic nonlinearity $g_2|\psi|^4\psi$ to Eq.\ (\ref{Eq_GPE}), where $g_2=24\ln\left(4/3 \right) N^2\hbar\omega_ra_s^2$, yields a 1D equation which approximates the integrability-broken nature of the full 3D GPE \cite{Khaykovich:2006}. We use the GPE equation modified with this quintic term to verify the regimes of validity of our results based on Eq.\ (\ref{Eq_GPE}).

In previous work \cite{Scharf:1992,Martin:2007, Martin:2008}, the following soliton-soliton interaction potential was employed to model solitons as individual particles:
\begin{equation}
V = -2\eta_1 \eta_2 \left(\eta_1 + \eta_2\right) \mathrm{sech}^2\left[\frac{2\eta_1\eta_2}{\eta_1+\eta_2}\tilde{q} \right],\label{Eq_vrel}
\end{equation}
where 
$\tilde{q}=m\left| g_{1D} \right|N q /\hbar^2$
is the dimensionless form of the relative position coordinate, $q$, of the solitons, and $\eta_i$ are effective soliton mass parameters, which take the value $\eta=1/8$ for two solitons of equal size  \cite{Martin:2007, Martin:2008}. This potential produces the exact position shifts for solitons emerging from collisions within Eq.\ (\ref{Eq_GPE}) when the axial potential $\omega=0$. The interaction potential was shown numerically also to provide a good description when combined with an external trapping potential in regimes where the solitons were well-separated between collisions, particularly when the solitons collided in-phase \cite{Martin:2008}.  

Within the particle model, the oscillation frequency of the solitons' relative position dynamics is given by %$\Omega = 2\pi/T$, where
%the period is 
\begin{equation}
\Omega = 2\pi/ \int_{-\tilde{q}_0}^{\tilde{q}_0}\frac{dx}{\sqrt{\left(\omega/4\right)\left(\tilde{q}_0^2 - x^2\right) - 4\eta^2\left(\mathrm{sech}^2\left(\eta \tilde{q}_0\right) - \mathrm{sech}^2\left(\eta x\right)\right)}}. \label{Eq_T}
\end{equation}
Here, $\pm \tilde{q}_0$ are the turning points of the dimensionless relative position coordinate, $\tilde{q}$. For solitons initially at rest (as in the recent experiment \cite{Nguyen:2014} and all cases analysed in this paper), $q_0$ is equal to the initial soliton separation. The frequency-shift may then be written $\Delta \omega = \Omega - \omega$.
We have found analytical solutions to Eq.\ (\ref{Eq_T}), in cases where the solitons are initially close or far apart.
In the limit $q_0 m|g_{1D}|N/\hbar\ll1$, i.e., where the solitons are at all times extremely close, we find:
\begin{equation}
\frac{\Delta \omega}{\omega} =  \frac{\sqrt{\omega^2 + 16 \left|g_{1D}\right|^4N^4\eta^4m^2/\hbar^6}}{\omega}. \label{Eq_smallq}
\end{equation}
For large initial separation, where $q_0 m|g_{1D}|N/\hbar\gg1$ and $ {\hbar^2} \ln\left(1 + {16\eta^2|g_{1D}|^2N^{2}/\hbar^{2}}{\omega^2q_0^{2}} \right) /{2\pi q_0\eta m|g_{1D}|N}  \ll 1$, we can assume that the position-shifts generated by Eq.\ (\ref{Eq_vrel}) occur on a timescale much faster than the trap period, and the oscillation period is found by merely adding the time-shifts generated by  Eq.\ (\ref{Eq_vrel}) during two collisions to the trap period, yielding:
\begin{equation}
\frac{\Delta \omega}{\omega} = \frac{-\chi}{1+\chi} , \label{Eq_largeq}
\end{equation}
where
\begin{equation}
\chi = \frac{-2\hbar^2}{q_0\eta m|g_{1D}|N}\ln\left(1 + \frac{16\eta^2|g_{1D}|^2N^{2}/\hbar^{2}}{\omega^2q_0^{2}} \right).
\end{equation}
%Note that in the limit that $q_0 m|g_{1D}|N/\hbar\rightarrow \infty$, it follows from Eq.\ (\ref{Eq_largeq}) that $\Delta \omega/\omega\rightarrow 0$.

\begin{figure}
\includegraphics[width=\columnwidth]{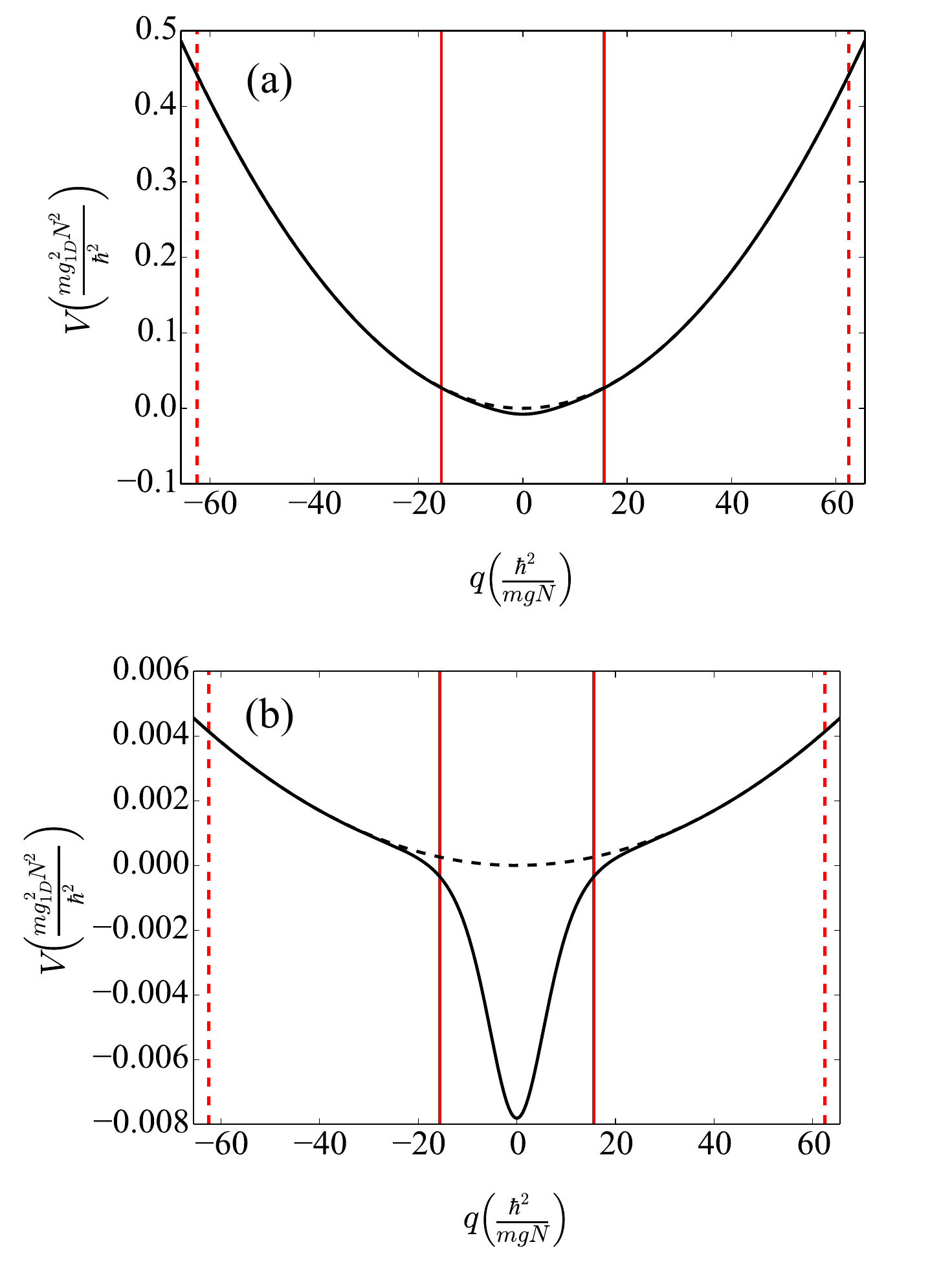}
\caption{(Colour online.) Effective potential for the system of two solitons in a harmonic trap, with scattering length $a_s =  -0.57 a_0$ and $N=56000$, and the mass parameter corresponding to the Lithium-7 atomic species. The dashed curved shows the harmonic trapping potential alone. The trap frequencies are (a) $\omega = 2\pi\times 31$Hz, and (b) $\omega = 2\pi\times 3$Hz. The solid (red) vertical lines are at $q_0=26\mu$m, and the dashed (red) vertical lines at  $q_0 = 106\mu$m.
}\label{Fig:vsol}
\end{figure}

The behaviour of the frequency shift between these two limits can be explained qualitatively by considering the effective potential, illustrated in Fig.\ \ref{Fig:vsol}. %When the harmonic trap strength is small, the bi-modal appearance of the potential becomes most obvious [Fig.\ \ref{Fig:vsol}(b)].
The potential has a clear bi-modal character, with a narrower interaction mode within the harmonic potential. In the limit where $q_0 m|g_{1D}|N/\hbar\ll1$, the solitons are extremely close with respect to the width of the inter-soliton potential (which becomes large when the solitons' interaction strength decreases). The dynamics are completely described by small harmonic oscillations, which increase in frequency as the interaction strength increases. 
When the interactions are increased further, or the solitons' initial separations are increased, the solitons may still be strongly bound within the interaction potential, but this potential is no longer effectively harmonic, and the frequency increase with interaction strength is less rapid. For even stronger interactions or wider separations, the inter-soliton potential becomes narrow compared with the initial soliton separation, and the solitons effectively `escape' the inter-soliton potential each collision. In this regime, the frequency shift levels off, and starts to decrease as the width of the inter-soliton potential vanishes in comparison with $q_0$. 

We provide for comparison the theoretical approximation derived in Ref.\ \cite{Nguyen:2014} based on the interaction Hamiltonian of a linear superposition of two Gaussian wavepackets:
\begin{equation}
\frac{\Delta\omega}{\omega} = \frac{-g_{1D}Na_{x}^2}{2\pi q_0^3\hbar\omega},\label{Eq_Malomed}
\end{equation}
where $a_x$ is the axial harmonic length of the trapping potential.
As opposed to the behaviour of the dynamics in the particle model, within Eq.\ (\ref{Eq_Malomed}), the shift is linear  in the interaction strength. The most marked difference between the prediction of the particle model and that of  Eq.\ (\ref{Eq_Malomed}) is that the frequency shift in Eq.\ (\ref{Eq_Malomed}) diverges to infinity, rather than to zero, as the interaction strength goes to infinity. In the opposite limit the particle model goes to zero quadratically, rather than linearly. However, the important comparison is for experimentally accessible regions ($N_s/N_c>-1$), and we are most interested in regimes where the difference between the predictions is large enough to resolve in experiment. In Ref.\ \cite{Nguyen:2014}, the initial soliton separation, $q_0 = 26\mu$m; number of atoms per soliton, $N_s=N/2=28000$; and radial and axial trap angular frequencies, $\omega_r = 2\pi\times 254$Hz and  $\omega = 2\pi\times31$Hz. The scattering length $a_s$ was varied between runs such that $N_s/N_c$ took values between $-0.53$ and $+0.55$. Note that in the regimes of positive $N_s/N_c$, the wavepackets are not strictly solitons, and such regimes are not considered in this paper. %We consider the experimental regime below, along with other experimentally-accessible regimes.

\begin{figure*}
%\centering
\includegraphics[width=2\columnwidth]{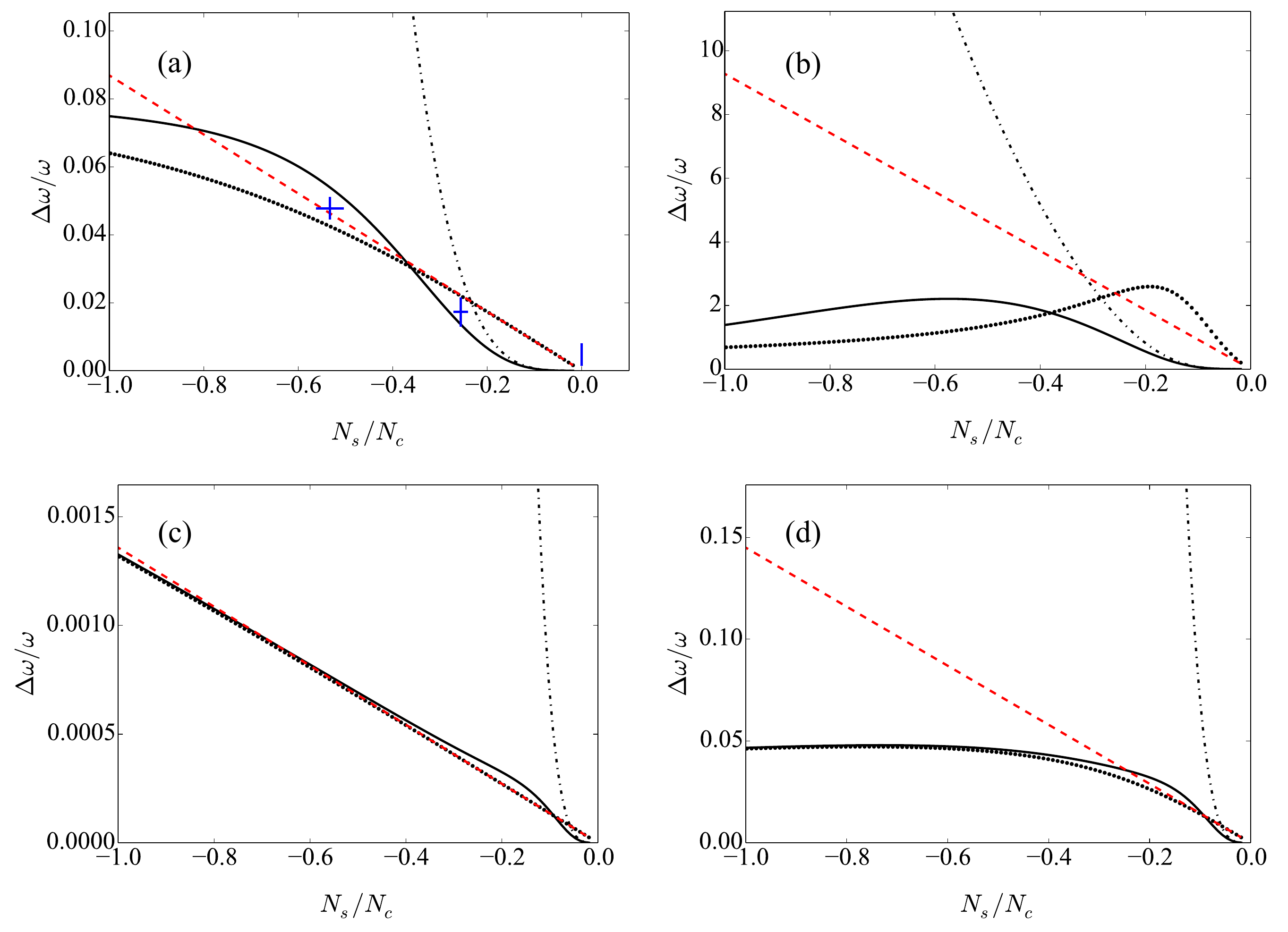}
\caption{(Colour online.) Relative frequency shifts versus $N_s/N_c$ for (a), (c) $\omega = 2\pi\times31$Hz, and (b), (d) $\omega = 2\pi\times 3$Hz. Plots (a), (b) are for initial soliton separation $q_0=26\mu$m, and (c), (d) are for $q_0 = 106\mu$m. The solid lines were determined by numerical evaluations of Eq.\ (\ref{Eq_T}). The dot-dashed lines are the asymptotic limit of small separations and weak interactions, given by Eq.\ (\ref{Eq_smallq}), the dotted lines are the asymptotic limit of large separations and strong interactions given by Eq. (\ref{Eq_largeq}). The dashed (red) lines give the approximate result from Eq.\ (\ref{Eq_Malomed}) and (blue) bars give the experimental results \cite{Nguyen:2014}.
} \label{Fig:dw}
\end{figure*}

Figure \ref{Fig:dw} shows the frequency shift from numerical evaluation of Eq.\ (\ref{Eq_T}), along with the curves for Eqs.\ (\ref{Eq_smallq}) and (\ref{Eq_largeq}) representing the limiting behaviour, and the theoretical approximation provided in Ref.\ \cite{Nguyen:2014}.
The relevant region of the experimental parameters lies within Fig.\ \ref{Fig:dw}(a); while Figs.\ \ref{Fig:dw}(b)-(d) show some other regimes of interest -- with different initial soliton separations or trap frequencies than in the experiment. 
The main deviation between the particle model and Eq.\ (\ref{Eq_Malomed}) is obvious in the large negative $N_s/N_c$ limit, where the frequency in the particle model levels off and then starts to decrease. This discrepancy is particularly apparent in Figs.\ \ref{Fig:dw}(b) and (d), where the axial trapping frequency is a factor of approximately 10 less than in the recent experiment. However, for the parameters of recent experiment [Fig.\ \ref{Fig:dw}(a)] the difference between the curves is probably within experimental error;  for large initial soliton separation and tight trapping [Fig.\ \ref{Fig:dw}(c)], the agreement between the curves is extremely close. 

%[INTEGRATE THIS WITH NEXT 2 PARAS]
%\section{Results}
%\subsection{Frequency-shift predictions}
%\subsection{Particle model validation}
Before issuing confident experimental predictions from the above particle model results, there are two reasons to verify the model's performance. Firstly, the particle model (and also the theoretical approximation of Ref.\ \cite{Nguyen:2014}) assumes that the solitons separate between collisions, and approach at a sufficient speed such that the relative-phase between the solitons plays no effect. It is clear from the results of Ref.\ \cite{Martin:2008} that for solitons slow enough that the collision time approaches the order of the trap period, out-of-phase collisions will have smaller frequency-shifts than those predicted by the particle model, whereas in-phase collisions are still well described. In the regimes where the initial separation was $q_0=26\mu$m \cite{Nguyen:2014}, the ansatz used to derive the particle model consists of initially slightly overlapping solitons, even for the narrowest solitons considered in the experiment. This obviously contradicts the assumption that the solitons always separate completely between collisions. Secondly, in some regimes, 3D effects to lead to condensate collapse during collisions, especially for $N_s/N_c>-0.5$ and in-phase solitons \cite{Nguyen:2014, Parker:2008}.

\begin{figure}
\includegraphics[width=\columnwidth]{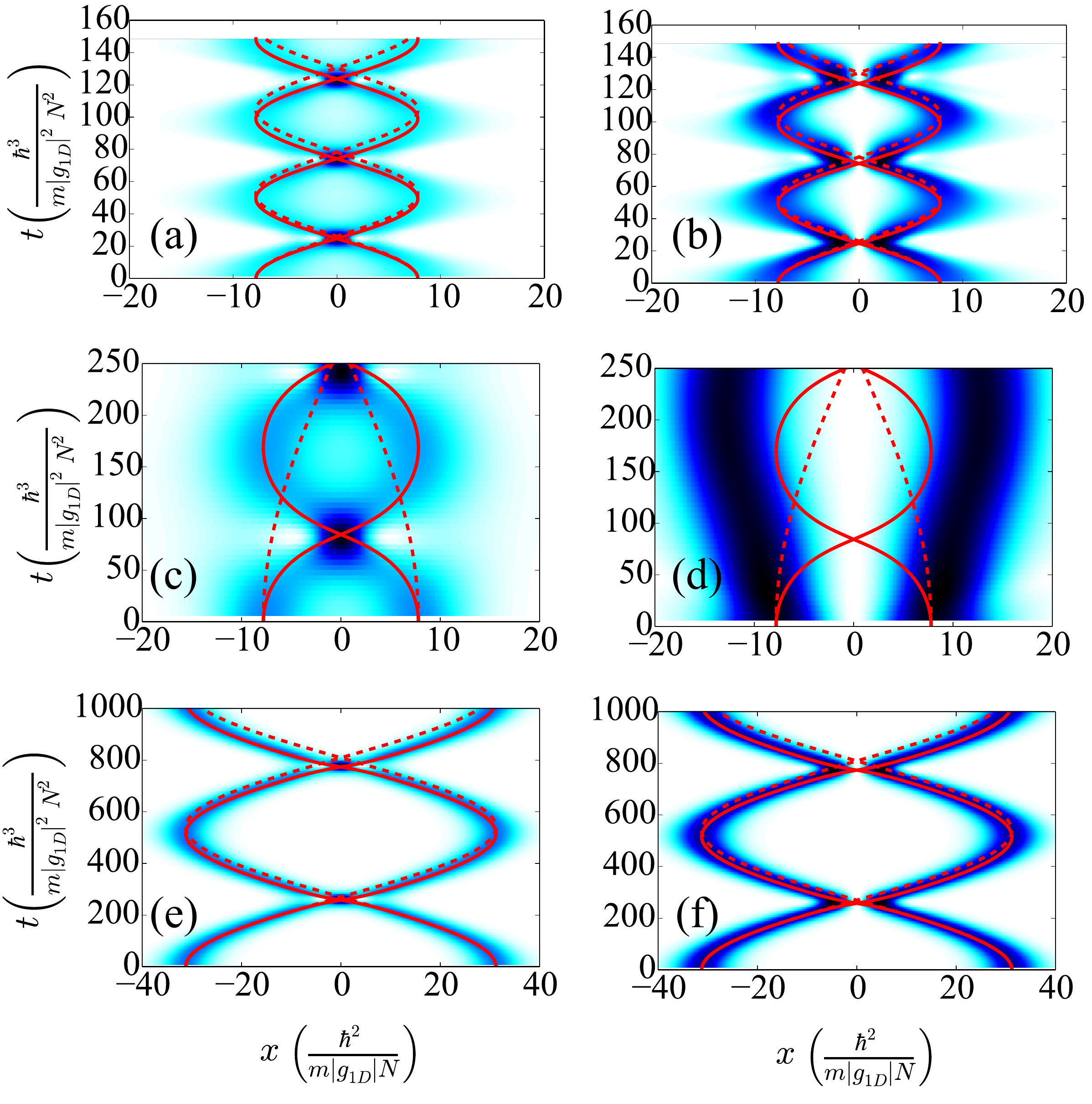}
\caption{(Colour online.) Atomic density within the GPE for initial separation: (a)-(d) $q_0=26\mu$m, (e)-(f) $q_0=106\mu$m; axial trap angular frequency: (a)-(b)  $\omega=2\pi\times 31$Hz, (c)-(f) $\omega=2\pi\times 3$Hz. The relative phase between the solitons is (a), (c), (e) $\phi=0$, (b), (d), (f) $\phi=\pi$. The scattering length $a_s=-0.57a_0$, the mass is that of the Lithium-7 atomic species and $N=56000$ such that $N_s/Nc=-0.53$. Trajectories within the particle model are given by full (red) lines, and the trajectories of non-interacting particles are given by dashed (red) lines for comparison. 
}\label{Fig:dens1d}
\end{figure}

We first evaluate on the phase dependence of the position shifts by integrating the GPE without the quintic term. Fig.\ \ref{Fig:dens1d}(a)-(b) shows GPE simulations two solitons for the experimental parameters with $N_s/N_c=-0.53$ for in- and out-of-phase collisions. The in-phase collisions are well described by the particle model, and, as expected, the predicted frequency shift is too large for the out-of-phase collisions. 
%Below we explore which regimes in Fig.\ \ref{Fig:dw} that are well described by the particle model, and for which relative phases, by using GPE simulations with and without the quintic term to simulate 3D effects.
Interestingly, we find that when the axial trap is weakened to $\omega = 2\pi\times 3$Hz [Fig.\ \ref{Fig:dens1d}(c)-(d)], the in-phase solitons form a bound state (soliton molecule) which is surprisingly well-described by the particle model (until it eventually de-phases at longer timescales). The out-of-phase solitons (which seem to repel each other) are badly described by the particle model.  Similar solitons with relative phases other than 0 and $\pi$ will have intermediate behaviour. We expect that the theoretical model employed in Ref.\ \cite{Khawaja:2011} would provide a better description of the GPE dynamics of these soliton molecules if it could be extended to incorporate the (weak) external trapping potential.

For larger initial soliton separations than used in the recent experiment, the particle model agrees very well with the GPE for any value of the relative phase. We illustrate this by simulating the system for an initial separation of four times that of recent experiment: $q_0 = 106\mu$m. For the experimental trap frequency, the frequency shift is too small to detect practically. We instead consider a weaker trap frequency of $\omega = 2\pi\times 3$Hz, resulting in slower solitons with easily observable frequency shifts. The agreement between the model and the GPE is illustrated in Fig.\ \ref{Fig:dens1d}(e)-(f). Note that it is this regime in which there is greatest improvement of the particle model over Eq.\ (\ref{Eq_Malomed}).

\begin{figure}
\includegraphics[width=\columnwidth]{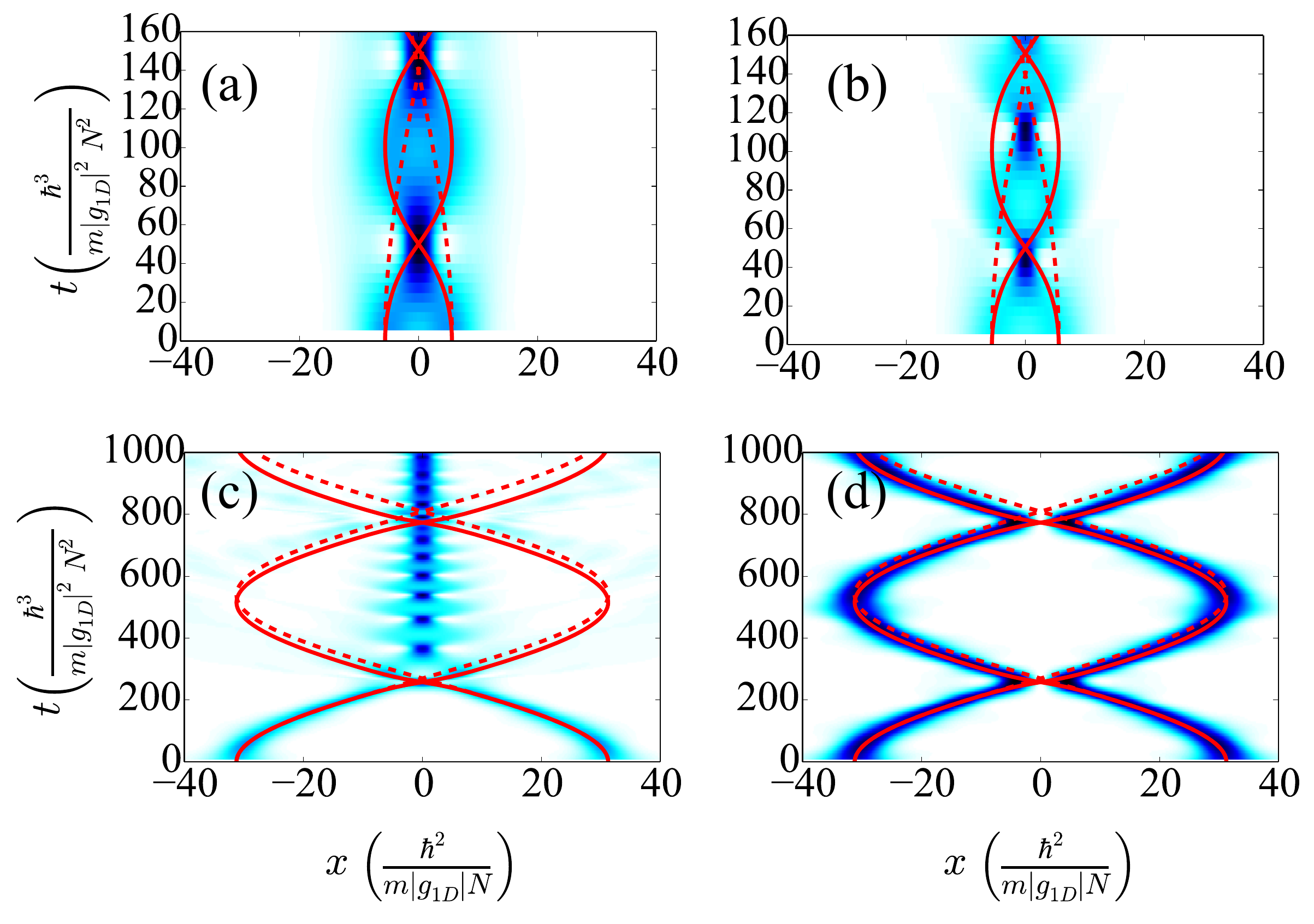}
\caption{(Colour online.) (a)-(b) Atomic density within the GPE for a bound state without (a) and with (b) a quintic nonlinearity simulating the 3D nature of the BEC. Here, $N_s/N_c=-0.41$, $\omega=2/pi\times 3$Hz. (c)-(d) Atomic density within the GPE for a bound state with a quintic nonlinearity for in-phase (c) and out-of-phase (d) collisions. Trajectories within the particle model are given by full (red) lines, and the trajectories of non-interacting particles are given by dashed (red) lines. 
}\label{Fig:dens3d}
\end{figure}

%\subsection{Collapse and 3D effects}
We now investigate 3D effects by integrating the 1D GPE with the quintic nonlinearity. We find that for interaction strength $N_s/N_c=-0.53$, the in-phase collisions are unstable, to collapse, but not the out-of-phase collisions [see Fig.\ \ref{Fig:dens3d}(c)-(d)]. This suggests that the in-phase collisions of Fig.\ \ref{Fig:dens1d}(a),(c) and (e) would not be realisable. However, slightly reducing the interaction strength to $N_s/N_c=-0.41$ produced qualitatively similar states which do not collapse. We find that the quintic term increases the oscillation frequency by a factor of approximately 1.5 [see figure 4(a)-(b)]. The frequency shifts of the unbound states remain unaffected [see Fig.\ \ref{Fig:dens3d}(d)], and it is here that the particle model is most useful. 

%\section{Conclusions}
In conclusion, we demonstrated the existence of regimes described better by the particle model than the theoretical approximation \cite{Nguyen:2014}, i.e., those of weak axial trap frequency and stronger interactions. We propose the extension of the recent experiment \cite{Nguyen:2014} to verify the particle model predictions. We also suggest that soliton molecules can be created if greater control of the soliton phase is possible, e.g., by applying a light-sheet potential to half of the condensate \cite{Burger:1999, Denschlag:2000, Becker:2008}. However, such states are highly dependent on relative-phase, and also on 3D effects. In future work we propose extending the formalism of \cite{Khawaja:2011} to include the effects of a harmonic trapping potential and 3D effects in order to describe soliton molecules.

We thank D. Schumayer and P. B. Blakie for useful discussions, and P. B. Blakie for assistance with preparation of Fig,\ \ref{Fig:dw}(a). 

\bibliography{Paper1}

\end{document}